\documentclass[aps,pra,twocolumn,showpacs,superscriptaddress]{revtex4} 
\usepackage{graphicx}
\usepackage{amsmath}
\usepackage{amssymb}
\usepackage{epstopdf}
\usepackage{amsfonts}
\usepackage{dcolumn}
\begin{document}
\title{Molecular branch of a small highly-elongated Fermi gas with an impurity:
Full three-dimensional versus effective one-dimensional description}
\author{Seyed Ebrahim Gharashi}
\affiliation{Department of Physics and Astronomy,
Washington State University,
  Pullman, Washington 99164-2814, USA}
\author{X. Y. Yin}
\affiliation{Department of Physics and Astronomy,
Washington State University,
  Pullman, Washington 99164-2814, USA}
\author{D. Blume}
\affiliation{Department of Physics and Astronomy,
Washington State University,
  Pullman, Washington 99164-2814, USA}
\date{\today}

\begin{abstract}
We consider an impurity immersed in a small Fermi gas under highly-elongated
harmonic confinement. The impurity interacts with the atoms of the
Fermi gas through an isotropic short-range potential with 
three-dimensional free-space $s$-wave scattering
length $a_{3\text{d}}$. We investigate the energies of the molecular branch,
i.e., the energies of the state that corresponds to a 
gas consisting of a weakly-bound diatomic molecule and ``unpaired'' atoms, 
as a function of the $s$-wave scattering length $a_{3\text{d}}$
and the ratio $\eta$ between 
the angular trapping frequencies in the tight and weak confinement
directions.
The energies obtained from our three-dimensional description that accounts for the
dynamics in the weak and tight confinement directions are compared with those
obtained within an effective one-dimensional framework, which
accounts for the dynamics in the tight confinement direction
via 
a renormalized one-dimensional coupling constant. 
Our theoretical results are related to recent experimental measurements.
\end{abstract}
\pacs{}
\maketitle

\section{Introduction}
\label{sec_introduction}
Ultracold few-atom systems provide clean model systems in which the system
parameters such as the interaction strength and confinement geometry
can be controlled with high 
accuracy~\cite{BlumeReview, GiorginiReview, BlochReview, ChinReview}.
Recently, two-component Fermi gases consisting of lithium atoms in
two different hyperfine states have been prepared and probed 
experimentally in highly elongated, nearly harmonic external traps with
an aspect ratio $\eta$ around 
ten~\cite{Jochim1,Jochim2,sala13,Jochim3,Jochim4,thesisGerhard}.
In two-component $^{6}$Li mixtures, 
the interspecies $s$-wave scattering length $a_{3\text{d}}$
can be tuned
by varying an external magnetic 
field in the vicinity of a Fano-Feshbach resonance~\cite{ChinReview}.
The intraspecies interactions are, to a very good approximation, negligible 
since the experiments operate at field strengths
that are quite far away from $p$-wave and higher partial wave resonances.
Thus, since the harmonic oscillator lengths $a_{\rho}$ and $a_z$ that
characterize the confinement in the tight and weak confinement
directions are much larger than the van der Waals length 
$r_{\text{vdW}}$ of the lithium-lithium
potential, the system dynamics is, to a very good approximation, governed
by the $s$-wave scattering length $a_{3\text{d}}$, the aspect ratio $\eta$
and the numbers $N_1$ and $N_2$ of fermions in the 
first and the second component, respectively.

This paper investigates the energetics of a single impurity immersed in a 
gas of fermions, referred to as the $(N-1,1)$ system, where $N$ denotes the total
number of particles. We focus on small systems with $N=2-4$ and 
determine the energy of the molecular branch. This is the branch 
that can be populated by preparing the system in the non-interacting regime,
i.e., with vanishing interspecies
$s$-wave scattering length, and by adiabatically
tuning the external magnetic field such that the $s$-wave scattering length 
takes small negative to infinitely large to positive values.
Since equal-mass two-component Fermi gases with short-range 
$s$-wave interactions do
only support weakly-bound dimers and not weakly-bound trimers
or tetramers~\cite{BlumeReview}, 
the molecular branch consists, roughly speaking, of a 
diatomic molecule that interacts with the ``unpaired'' atoms.
This work compares the energies obtained from a full three-dimensional treatment
and an effective one-dimensional treatment. The latter is
expected to break down when the size of the dimer becomes comparable to the
harmonic oscillator length
$a_{\rho}$ of the tight confinement direction. This is the regime
where the system is expected to ``explicitly feel'' the degrees of freedom
associated with the tight confinement direction. 
Assuming harmonic confinement, we provide 
quantitative comparisons of the full three-dimensional and approximate
one-dimensional treatments 
as a function of the $s$-wave scattering length and
the aspect ratio $\eta$.
We compare our energies of the molecular branch 
to those determined experimentally
via radio frequency spectroscopy~\cite{Jochim3,thesisGerhard}.
While the dynamics of the molecular branch of the
$(1,1)$ system has been investigated extensively in the 
literature~\cite{bolda03,Idziaszek05,Idziaszek06,sala13},
that of the $(2,1)$ and $(3,1)$ systems has not
yet been investigated comprehensively theoretically
although first steps have been taken~\cite{mora05,mora05a}.

A full quantitative theoretical  understanding of the 
energetics of the molecular
branch of small Fermi gases under highly-elongated 
confinement is important for several reasons.
Experiments on few-fermion systems 
are becoming more and more precise~\cite{Jochim5,sala13}, 
opening the door for 
quantitative studies of a myriad of few-atom phenomena.
Moreover, a thorough understanding of few-body systems serves as a guide
for larger systems where the analysis relies, in many cases, on 
approximate treatments and plays a crucial ingredient in mapping out
the transition from few- to many-body 
physics~\cite{Jochim4,Jochim3,thesisGerhard}.

The remainder of this paper is organized as follows. 
Section~\ref{sec_hamiltonian} introduces the system Hamiltonian 
and reviews briefly how to solve the corresponding time-independent 
Schr\"odinger equation. Section~\ref{sec_results} presents our results 
for the full three-dimensional
and the approximate one-dimensional treatments. 
Lastly, Sec.~\ref{sec_summary}
summarizes.

\section{System Hamiltonian}
\label{sec_hamiltonian}
The three-dimensional system Hamiltonian $H_{3\text{d}}$ for the $N$
fermions of
mass $m$ reads
\begin{eqnarray}
\label{eq_ham3d}
H_{3\text{d}} = \sum_{j=1}^N 
\left[ \frac{-\hbar^2}{2m} \nabla_{\vec{r}_j}^2
+ \frac{1}{2} m \left( \omega_{\rho}^2 \rho_j^2 + \omega_z^2 z_j^2 \right) 
\right] + \nonumber \\
\sum_{j=1}^{N_1} \sum_{k=N_1+1}^N V_{\text{tb}}(\vec{r}_{jk}),
\end{eqnarray}
where $\vec{r}_j$ denotes the position vector of the
$j$th atom measured with respect to the trap center, $\vec{r}_j=(x_j,y_j,z_j)$,
and $\rho_j^2=x_j^2+y_j^2$.
The angular trapping frequencies $\omega_{\rho}$ and $\omega_z$
are related through the aspect ratio $\eta$,
where $\eta=\omega_{\rho}/\omega_z$.
Throughout, we consider 
elongated confining geometries with $\eta>1$ ($\eta=2,3,\cdots$).
The interaction potential $V_{\text{tb}}(\vec{r}_{jk})$ between the unlike
fermions is modeled by two different potentials, the regularized
zero-range Fermi-Huang pseudopotential 
$V_{\text{ps}}(\vec{r}_{jk})$~\cite{ferm34,Huang57,HuangText},
\begin{eqnarray}
\label{eq_pseudo}
V_{\text{ps}}(\vec{r}_{jk}) = \frac{4 \pi \hbar^2 a_{3\text{d}}}{m} \delta^{(3)}(\vec{r}_{jk})
\frac{\partial}{\partial r_{jk}} r_{jk},
\end{eqnarray}
and a finite-range Gaussian potential $V_{\text{g}}(\vec{r}_{jk})$ with depth $V_0$
($V_0 \ge 0$) and range $r_0$,
\begin{eqnarray}
\label{eq_gaussian}
V_{\text{g}}(\vec{r}_{jk}) = -V_0 \exp \left[-
\left( \frac{r_{jk}}{\sqrt{2}r_0} \right)^2 \right],
\end{eqnarray}
where $\vec{r}_{jk}=\vec{r}_j-\vec{r}_k$ and $r_{jk}=|\vec{r}_{jk}|$.
In Eq.~(\ref{eq_pseudo}), $a_{3\text{d}}$ denotes the three-dimensional free-space 
zero-energy $s$-wave scattering length.
To compare the eigenenergies of $H_{3\text{d}}$ for $V_{\text{ps}}$ and $V_{\text{g}}$,
the depth $V_0$ and range $r_0$ are adjusted such that the free-space scattering lengths
of the two potentials agree. Throughout, we restrict ourselves to the
regime where $V_{\text{g}}$ supports zero free-space $s$-wave bound states for
$a_{3\text{d}}<0$ and one free-space $s$-wave bound state for $a_{3\text{d}}>0$.
We are interested in the regime where the range $r_0$ is much smaller than the
oscillator lengths $a_{\rho}$ and $a_z$, 
where $a_{\rho}=\sqrt{\hbar/(m \omega_{\rho})}$
and 
$a_{z}=\sqrt{\hbar/(m \omega_{z})}$.

For sufficiently large $\eta$, the low-energy
physics
described by
the three-dimensional Hamiltonian $H_{3\text{d}}$ is expected to be reproduced by the
effective one-dimensional Hamiltonian $H_{1\text{d}}$~\cite{Olsh98},
\begin{eqnarray}
\label{eq_ham1d}
H_{1\text{d}} = 
\sum_{j=1}^N \left(
\frac{-\hbar^2}{2m} \frac{\partial^2}{\partial z_j^2} 
+
\frac{1}{2} m \omega_z^2 z_j^2 \right) + \nonumber \\
\sum_{j=1}^{N_1} \sum_{k=N_1+1}^N g_{1\text{d}} \delta^{(1)}(z_{jk}) + N \hbar \omega_{\rho},
\end{eqnarray}
where $z_{jk}=z_j-z_k$. The effective one-dimensional 
coupling constant 
$g_{1\text{d}}$~\cite{Olsh98},
\begin{eqnarray}
\label{eq_g1d}
\frac{g_{1\text{d}}}{\hbar \omega_{\rho} \; a_{\rho}} = \frac{2 a_{3\text{d}}}{a_{\rho}}
\left( 1 + \frac{\zeta(1/2)}{\sqrt{2}} \frac{a_{3\text{d}}}{a_{\rho}} \right) ^{-1}
\end{eqnarray}
with $\zeta(1/2) \approx -1.46035$,
 has been derived by analyzing $H_{3\text{d}}$ with $\omega_z=0$
for two particles
with zero-range interactions.
The renormalization of the one-dimensional coupling constant,
i.e., the second term in the round brackets 
on the right hand side of Eq.~(\ref{eq_g1d}),
 accounts for the occupation of excited transverse modes during the
collision process (i.e., for virtual excitations).
An improved description is obtained if 
the effective one-dimensional coupling constant 
is made to depend explicitly on the energy~\cite{olshanii2,granger}
(see Sec.~\ref{sec_results}
for more details). 
It has been shown that the eigenenergies of the lowest gas-like state
of $H_{1\text{d}}$
are in good agreement with those of $H_{3\text{d}}$ if the aspect ratio 
is sufficiently large~\cite{Idziaszek06,Gharashi12,debraj}. 
More specifically, the energy spacing $\hbar \omega_{\rho}$
in the
tight confinement direction has to be larger than $(N_1-1)\hbar \omega_z$.
This paper investigates the 
properties of 
the molecular
branch
for $H_{3\text{d}}$ and $H_{1\text{d}}$, 
i.e., we primarily
focus on the regime where the eigenenergies are smaller
than the ground state energy of the non-interacting Hamiltonian.

To determine the eigenstates and eigenenergies of $H_{3\text{d}}$ and 
$H_{1\text{d}}$, we separate the center of mass and relative degrees of freedom.
The relative Schr\"odinger equation
for the one-dimensional $(1,1)$, $(2,1)$ and $(3,1)$ systems is solved
using the approaches introduced in Ref.~\cite{Busch98},
Ref.~\cite{Gharashi12} and Ref.~\cite{Gharashi13}, respectively.
The relative Schr\"odinger equation for the three-dimensional 
$(1,1)$ and $(2,1)$ systems 
interacting through $V_{\text{ps}}$ is solved using the approaches introduced
in Ref.~\cite{Idziaszek06} and Ref.~\cite{Gharashi12}, respectively.
 For the $(3,1)$ system, we only consider the finite-range interaction model
$V_g$.
The relative Schr\"odinger equation
for the three-dimensional $(1,1)$ system with finite-range 
interactions is solved using a B-spline approach
while that for the three-dimensional $(2,1)$ and $(3,1)$ 
systems is solved using an explicitly
correlated basis set expansion approach~\cite{CGReview, CGbook}.
For a fixed $a_{3\text{d}}$,
we perform calculations for several $r_0$ and then extrapolate to the zero-range limit.
Since the center of mass motion is uneffected by the 
two-body interactions, 
Sec.~\ref{sec_results}
reports
the relative eigenenergies $E_{N_1,N_2}^{\text{3d}}$ and
$E_{N_1,N_2}^{\text{1d}}$.

\section{Results}
\label{sec_results}
\begin{figure}
\centering
\includegraphics[angle=0,width=0.4\textwidth]{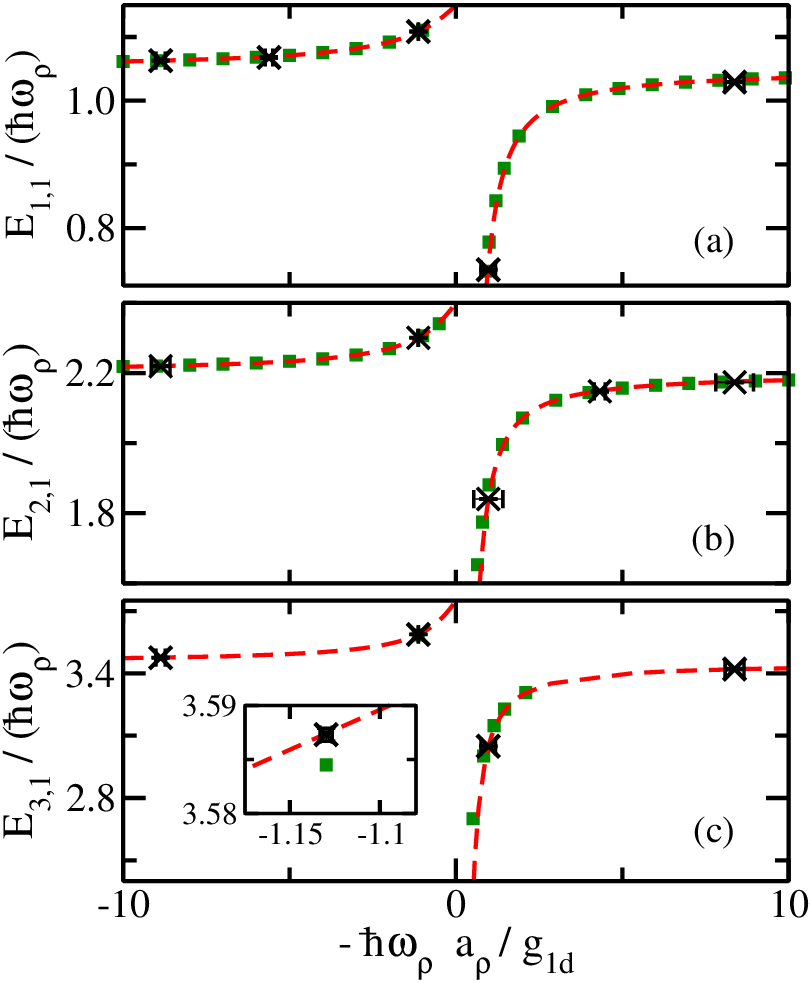}
\caption{(Color online)
Relative zero-range energies $E_{N_1,N_2}$
as a function of $-1/g_{1\text{d}}$ for $\eta=10$.
Squares and 
dashed lines show the relative energies 
obtained from the full three-dimensional treatment and the effective
one-dimensional treatment, respectively, for 
(a) the $(1,1)$ system,  
(b) the $(2,1)$ system, and  
(c) the $(3,1)$ system.
Crosses with errorbars show experimental results 
(the experimental data points are taken from Figure~A.21. of
Ref.~\cite{thesisGerhard}).
The inset in panel~(c) shows a blow-up of the energetically lowest lying
gas-like branch with negative $-g_{\text{1d}}$.
 }\label{fig_energyeta10}
\end{figure}

This section discusses the eigenenergies of the
$(1,1)$, $(2,1)$ and $(3,1)$ systems.
Squares
in Fig.~\ref{fig_energyeta10} show the relative 
zero-range energies $E_{N_1,N_2}^{\text{3d}}$ of the energetically
lowest lying molecular branch and 
the energetically lowest lying gaslike state
as a function of $-1/g_{1\text{d}}$
for $\eta=10$. 
To make this plot, the three-dimensional scattering length $a_{3\text{d}}$
was converted to the effective one-dimensional coupling constant
$g_{1\text{d}}$ using Eq.~(\ref{eq_g1d}).
The non-interacting limit is reached for $-1/g_{1\text{d}}= \pm \infty$
and the infinitely strongly interacting regime for $|1/g_{1\text{d}}|=0$.
The points $-1/g_{1\text{d}}= \pm \infty$ correspond to $a_{3\text{d}}=\mp0$
while the point $|1/g_{1\text{d}}|=0$ corresponds to 
$a_{3\text{d}} = \sqrt{2} a_{\rho}/|\zeta(1/2)|$.
The eigenstates corresponding to
the relative eigenenergies of the $(1,1)$ system
shown in Fig.~\ref{fig_energyeta10}(a) are characterized
by an even parity in $z$, i.e., $\Pi_z=+1$, and vanishing 
projection quantum number $M$ and positive parity $\Pi_{\hat{\rho}}$
in the $xy$-plane~\cite{explain_parity}.
The eigenstates corresponding to
the relative eigenenergies of the $(2,1)$ and $(3,1)$ systems
shown in Figs.~\ref{fig_energyeta10}(b) and 
\ref{fig_energyeta10}(c), in contrast,
are characterized by $(\Pi_z,M,\Pi_{\hat{\rho}})=(-1,0,+1)$.
The negative parity in the $z$-direction is a consequence of the 
fact that the majority particles have to obey the
Pauli exclusion principle and that the single particle
harmonic oscillator states in one dimension have alternating 
even and odd parity (even parity for the principal quantum numbers
$n_z=0,2,\cdots$ and
odd parity for the principal quantum numbers
$n_z=1,3,\cdots$).

For comparison, dashed lines show the relative energies obtained 
within the one-dimensional framework.
The agreement between the three- and 
one-dimensional energies of the energetically lowest
lying gaslike branch is excellent (i.e., better
than about $0.5$\%) for
all interaction strengths considered. 
For the $(1,1)$ and $(2,1)$ systems, quantitative comparisons 
have been
presented in Refs.~\cite{Idziaszek06, Gharashi12}.
Figure~\ref{fig_energyeta10}(c) 
shows the three-dimensional energies of the energetically lowest lying
gas-like state of the $(3,1)$ system for one $g_{1\text{d}}$ value,
namely for $g_{1\text{d}}=0.8850 \hbar \omega_{\rho} a_{\rho}$
[see inset of Fig.~\ref{fig_energyeta10}(c)].
We choose this value since this is one of the 
coupling strengths at which the experiments of the Heidelberg group
were performed~\cite{Jochim3,thesisGerhard}.
Our three-dimensional zero-range 
energy agrees with the corresponding one-dimensional
energy to $0.1$\%. 
The eigenenergies deduced from radio frequency spectroscopy
measurements~\cite{thesisGerhard} are shown by crosses with errorbars.
The experimental data have been, using theoretical
three- and one-dimensional energies for the $(1,1)$ and $(2,1)$
systems~\cite{Jochim3,thesisGerhard},
converted
to one-dimensional energies and 
should be considered as 
aproximations to 
the relative
eigenenergies of $H_{1\text{d}}$.
The agreement between the theoretical results and the
energies deduced from experiment is at the few percent level,
underlining the precision and control of modern cold atom experiments and the
need for accurate theoretical treatments.

\begin{figure}
\vspace*{0.5in}
\centering
\includegraphics[angle=0,width=0.4\textwidth]{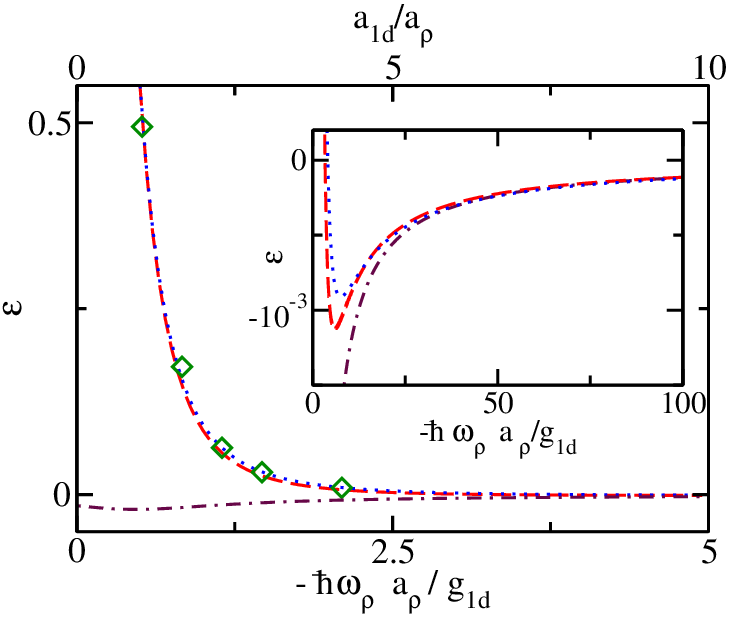}
\vspace*{0.5in}
\caption{(Color online)
Scaled interaction energy difference 
$\epsilon$ for the energetically lowest lying
molecular branch
as a function of $-1/g_{1\text{d}}$ for $\eta=10$.
The dashed line, dotted line
and diamonds show $\epsilon$
for the $(1,1)$, $(2,1)$ and $(3,1)$ systems,
respectively; in these calculations, the interaction
energy $E_{N-1,1}^{1\text{d},\text{int}}$ is calculated using
$g_{1\text{d}}$.
The dash-dotted line shows $\epsilon$ for the $(1,1)$ system
for the case where
the interaction
energy $E_{N-1,1}^{1\text{d},\text{int}}$ is calculated using
$g_{1\text{d}}^{\text{trap}}$, accounting for the energy-dependence
of $\zeta(1/2,1-\epsilon_{2\text{b}}/2)$ but neglecting the ``higher-order''
Hurwitz zeta functions. 
The axis label on top shows the one-dimensional scattering
length $a_{1\text{d}}$.
The inset shows $\epsilon$ for large $-1/g_{\text{1d}}$. 
 }\label{fig_energyratioeta10}
\end{figure}

For the energetically lowest lying molecular branch of the $(1,1)$, $(2,1)$
and $(3,1)$
systems, the agreement between the
three-dimensional energies
(squares) and the one-dimensional energies
(dashed lines) is excellent for large positive $-1/g_{1\text{d}}$ but
deteriorates as $-1/g_{1\text{d}}$ decreases 
(see also Fig.~\ref{fig_energyratioeta10}).
Qualitatively, this can be understood by realizing that the loosely
bound dimer is very large for  very 
positive $-1/g_{1\text{d}}$.
In fact, the size of the loosely bound dimer is directly proportional
to $|1/g_{1\text{d}}|$~\cite{Busch98,Olsh98,olshanii2}. 
As $-1/g_{1\text{d}}$ decreases and
approaches zero, the size of the dimer
decreases, implying that the system dynamics is no longer effectively
one-dimensional but that the dimer is sufficiently small to ``probe'' the
dynamics associated with the tight confinement direction.
Since three- and higher-body bound states are absent,
this qualitative argument applies not only to the $(1,1)$ system
but also to the $(2,1)$ and $(3,1)$ systems.
The experimental data for the molecular branch 
(see Fig.~\ref{fig_energyeta10}) are limited to fairly
large $-1/g_{1\text{d}}$ and 
the errorbars are not 
small enough to discriminate between the
three- and one-dimensional frameworks.

To quantify the applicability of the one-dimensional treatment for the
molecular branch, we define the scaled interaction energy
difference $\epsilon$,
\begin{eqnarray}
\epsilon = \frac{E_{N-1,1}^{3\text{d,int}}-E_{N-1,1}^{1\text{d,int}}}{|E_{N-1,1}^{3\text{d,int}}|},
\end{eqnarray}
where $E_{N-1,1}^{3\text{d,int}}$
is defined as the difference between the relative three-dimensional energy
$E_{N-1,1}^{\text{3d}}$ of the molecular branch and the non-interacting 
ground state energy, 
$E_{N-1,1}^{3\text{d,int}}=E_{N-1,1}^{\text{3d}}-E_{N-1,1}^{\text{NI}}$,
where $E_{N-1,1}^{\text{NI}}=(N-1)^2\hbar \omega_z/2+(N-1) \hbar \omega_{\rho}$.
The interaction energy 
$E_{N-1,1}^{1\text{d,int}}$ obtained from the one-dimensional treatment is defined
in an analogous way
(note that the non-interacting ground state
energies of $H_{3\text{d}}$ and $H_{1\text{d}}$
are identical).
Figure~\ref{fig_energyratioeta10} 
shows the quantity $\epsilon$
as a function
of $-1/g_{1\text{d}}$ for $\eta=10$.
To determine $\epsilon$, we use the three-dimensional
energies in the limit of zero-range interactions.
The axis label on the top shows the one-dimensional scattering length $a_{1\text{d}}$,
which
is related to the one-dimensional coupling constant $g_{1\text{d}}$ via
$a_{1\text{d}}=- 2 \hbar^2 / (m g_{1\text{d}})$.
Both $E_{N-1,1}^{3\text{d,int}}$ and $E_{N-1,1}^{1\text{d,int}}$ are negative.
The quantity $\epsilon$ is small and negative for large
$a_{\text{1d}}$ 
(see inset of Fig.~\ref{fig_energyratioeta10})
and changes sign around $a_{\text{1d}} \approx 5-10 a_{\rho}$.
The positive value of $\epsilon$ 
for $a_{\text{1d}} \lesssim 5-10 a_{\rho}$ indicates that the energy
obtained from the three-dimensional
treatment lies above the energy
obtained from the one-dimensional treatment.
When $a_{1\text{d}}$ is large, corresponding to a large dimer,
the molecular branch is well described by 
the effective one-dimensional Hamiltonian. However, as $a_{1\text{d}}$ 
decreases and approaches $a_{\rho}$, the 
effective one-dimensional treatment deteriorates
in a similar manner for the 
$(1,1)$, $(2,1)$ and $(3,1)$ systems.
At $a_{1\text{d}}/a_{\rho}=1$, e.g., $\epsilon$ is equal to 
$0.55$, $0.55$ and $0.51$
for the $(1,1)$, $(2,1)$ and $(3,1)$ systems, respectively.

To better
understand the behavior of $\epsilon$, we start with the
implicit eigenequation
for the three-dimensional $(1,1)$ system~\cite{Idziaszek05,Idziaszek06},
\begin{eqnarray}
\label{eq_eigen3d}
\mathcal{F}_{3\text{d}}(\epsilon_{2\text{b}},\eta) = 
-\frac{ \sqrt{2 \eta} a_{\rho}}{a_{3\text{d}}},
\end{eqnarray}
where
\begin{eqnarray}
\label{eq_f3d}
\mathcal{F}_{3\text{d}}(\epsilon_{2\text{b}},\eta) = 
\nonumber \\
\frac{1}{\sqrt{\pi}}
\int_0^\infty
\left(
\frac{\eta  e^{\eta \epsilon_{2\text{b}} t/2}}
{\sqrt{1- e^{-t}} (1- e^{-\eta t})}
-\frac{1}{t^{3/2}} 
\right) dt
\end{eqnarray}
and 
$\epsilon_{2\text{b}}=E_{1,1}^{3\text{d},\text{int}}/(\hbar \omega_{\rho})$.
Equation~(\ref{eq_f3d}) holds for $\epsilon_{2\text{b}}<0$
but can be extended to positive $\epsilon_{2\text{b}}$ through
analytic continuation~\cite{Idziaszek06}.
To derive an effective one-dimensional eigenequation,
we expand the integrand of Eq.~(\ref{eq_f3d}), assuming small $1/\eta$,
around $t=0$.
Using this expansion in Eq.~(\ref{eq_eigen3d}) yields
\begin{eqnarray}
\label{eq_eigen1d}
\frac{\Gamma(-\eta  \epsilon_{2\text{b}}/2)}
{\sqrt{2} \Gamma(-\eta \epsilon_{2\text{b}}/2 +1/2)} 
= -\frac{2 \hbar \omega_{\rho}  a_{\rho}}{ \sqrt{\eta}g_{1\text{d}}^{\text{trap}}},
\end{eqnarray}
where $g_{1\text{d}}^{\text{trap}}$ denotes 
the ``trap-corrected'' effective one-dimensional
coupling constant, 
\begin{widetext}
\begin{eqnarray}
\label{eq_g1dtrap}
\frac{g_{1\text{d}}^{\text{trap}}}{\hbar \omega_{\rho}  a_{\rho}} = 
\frac{2 a_{3\text{d}}}{a_{\rho}}
\left\{1 + 
\left[ 
\frac{\zeta(1/2, 1 - \epsilon_{2\text{b}}/2)}{\sqrt{2}} +
\frac{\zeta(3/2, 1 - \epsilon_{2\text{b}}/2)}{8 \sqrt{2} \; \eta} +
\frac{\zeta(5/2, 1 - \epsilon_{2\text{b}}/2)}{128 \sqrt{2} \; \eta^2} +
\dots
\right]
\frac{a_{3\text{d}}}{a_{\rho}}
\right\}^{-1},
\end{eqnarray}
\end{widetext}
and $\zeta(\cdot,\cdot)$ the Hurwitz zeta function.
If the series in Eq.~(\ref{eq_g1dtrap}) goes to sufficiently large order,
the low-energy part of the eigenspectrum
determined by solving Eq.~(\ref{eq_eigen1d}) 
agrees very well with that determined
by solving Eq.~(\ref{eq_eigen3d}).
It is important to note that 
Eq.~(\ref{eq_eigen1d}) is the implicit eigenequation one obtains by
solving the relative Schr\"odinger
equation for the trapped two-particle
system interacting through a zero-range potential with coupling constant 
$g_{1\text{d}}^{\text{trap}}$~\cite{Busch98}.
Comparing the effective
one-dimensional coupling constants
$g_{1\text{d}}^{\text{trap}}$ and $g_{1\text{d}}$, 
we notice two things.
First, 
the $\zeta(1/2)$ term in Eq.~(\ref{eq_g1d}) is replaced
by the energy-dependent Hurwitz zeta function. 
For small $|\epsilon_{2\text{b}}|$, the Hurwitz zeta
function can be Taylor exanded,
$\zeta(n/2,1-\epsilon_{2\text{b}}/2)
\approx \zeta(n/2)+\zeta(1+n/2)n \epsilon_{2\text{b}}/4+\cdots$,
showing that $\zeta(n/2,1-\epsilon_{2\text{b}}/2)$
reduces to $\zeta(1/2)$ for
$\epsilon_{2\text{b}}=0$.
Second, $g_{1\text{d}}^{\text{trap}}$ contains
corrections that are suppressed by increasing powers of
$1/\eta$ [the second and third terms in
square brackets on the right hand side of Eq.~(\ref{eq_g1dtrap})]
and thus vanish as $\eta \rightarrow \infty$.

To see how the energy-dependence and the higher-order
corrections in $1/\eta$ affect the energy spectrum
of the $(1,1)$ system,
we consider the weakly- and more strongly-interacting regimes separately.
Expanding Eq.~(\ref{eq_eigen1d}) around $\epsilon_{2\text{b}}=0$,
we find 
\begin{eqnarray}
\label{eq_smallg1dtrap}
\frac{E_{1,1}^{1\text{d},\text{int}}}{\hbar \omega_{\rho}} = 
c_1^{(1)}\frac{g_{1\text{d}}^{\text{trap}}}{\hbar \omega_{\rho}  a_{\rho}}+
c_1^{(2)} \left(\frac{g_{1\text{d}}^{\text{trap}}}{\hbar \omega_{\rho}  a_{\rho}} \right)^2+ \nonumber \\
c_1^{(3)} \left(\frac{g_{1\text{d}}^{\text{trap}}}{\hbar \omega_{\rho}  a_{\rho}} \right)^3
+ \dots
\end{eqnarray}
or, rewriting this expression in terms of $g_{1\text{d}}$,
\begin{eqnarray}
\label{eq_smallg1d}
\frac{E_{1,1}^{1\text{d},\text{int}}}{\hbar \omega_{\rho}} = 
c_1^{(1)}\frac{g_{1\text{d}}}{\hbar \omega_{\rho}  a_{\rho}} +
(c_1^{(2)} + c_2^{(2)})
\left(\frac{g_{1\text{d}}}{\hbar \omega_{\rho}  a_{\rho}} \right)^2 + 
\nonumber \\
(c_1^{(3)} + c_2^{(3)} + c_3^{(3)})
\left(\frac{g_{1\text{d}}}{\hbar \omega_{\rho}  a_{\rho}} \right)^3 
+ \dots.
\end{eqnarray}
The $c_k^{(j)}$-coefficients are listed in Table~\ref{tab1}.
\begin{table}
\caption{Coefficients entering into Eqs.~(\ref{eq_smallg1dtrap})
and (\ref{eq_smallg1d}).}
\begin{tabular}{lc}
$c_1^{(1)}$ & 
$\frac{1}{\sqrt{2\pi  \eta}}$
\\
\hline
$c_1^{(2)}$ & 
$-\frac{\ln(2)}{2\pi}$
\\
\hline
$c_1^{(3)}$ & 
$\sqrt{\frac{\eta}{2\pi}} 
\left(- \frac {\pi} {48} + \frac {3\ln(2)^2} {4\pi} \right)$ 
\\
\hline
$c_2^{(2)}$ & 
$ -\frac{c_1^{(1)}}{\sqrt{8}} 
\left( 
\frac{\zeta(3/2)}{2^3 \eta}
+ \frac{\zeta(5/2)}{2^7 \eta^{2}}
+ \dots 
\right)$
\\
\hline
$c_2^{(3)}$ & 
$ \frac{-c_1^{(2)}}{\sqrt{2}} 
\left( 
 \frac{\zeta(3/2)}{2^3  \eta}
+ \frac{\zeta(5/2)}{2^7 \eta^2}
+ \dots 
\right) + $ \\
& $
\frac{c_1^{(1)}}{8} 
\left( 
 \frac{\zeta(3/2)}{2^3  \eta}
+ \frac{\zeta(5/2)}{2^7 \eta^2}
+ \dots 
\right)^2$
\\ 
\hline
$c_3^{(3)}$ & 
$-\frac{1}{\sqrt{2}\pi} 
\left( 
\frac {\zeta(3/2)} {2^4 \eta} 
+\frac {3\zeta(5/2)} {2^7 \eta^2} 
+ \dots 
\right)$ 
\end{tabular}
\label{tab1}
\end{table}
The coefficients $c_{2}^{(2)}$ and $c_{2}^{(3)}$
arise from the energy-independent parts of the
$\zeta(n/2, 1-\epsilon_{2\text{b}}/2)$ terms
($n=3,5,\cdots$) 
in Eq.~(\ref{eq_g1dtrap})
while 
the coefficient $c_3^{(3)}$ arises from 
the leading-order energy-dependence
of the $\zeta(n/2, 1-\epsilon_{2\text{b}}/2)$ terms
($n=1,3,\cdots$).
This analysis shows that the trap corrections encapsulated by
$c_2^{(2)}$ dominate,
in the small $g_{1\text{d}}$ limit, over 
higher-order trap corrections and the 
energy-dependence of the Hurwitz zeta
function
encapsulated 
respectively by $c_3^{(2)}$ and $c_3^{(3)}$.
The coefficient $c_2^{(2)}$ is negative and neglecting it,
as done in our determination of $E_{1,1}^{1\text{d},\text{int}}$,
leads to a larger energy and thus to a negative
$\epsilon$ for the $(1,1)$ system in the weakly-interacting
regime
(see the dashed line in the inset of Fig.~\ref{fig_energyratioeta10}).

When $|\epsilon_{2\text{b}}|$ is not small compared to $1$,
the energy-dependence of the Hurwitz
zeta functions in Eq.~(\ref{eq_g1dtrap})
plays an important role. 
To demonstrate this,
the dash-dotted line in Fig.~\ref{fig_energyratioeta10}
shows the quantity $\epsilon$ for the $(1,1)$ system 
calculated accounting for the leading-order 
energy-dependence of $g_{1\text{d}}^{\text{trap}}$.
Specifically, the one-dimensional $(1,1)$ energy is calculated
using $g_{1\text{d}}^{\text{trap}}$ 
neglecting the second and third terms in the
square brackets in Eq.~(\ref{eq_g1dtrap}).
The fact that the dash-dotted line in Fig.~\ref{fig_energyratioeta10}
is close to zero for all $g_{1\text{d}}$ 
demonstrates that the leading-order energy
dependence yields the dominant correction
when $|g_{1\text{d}}|/(\hbar \omega_{\rho} a_{\rho})$ is appreciable
(i.e., not small compared to $1$).

As already pointed out earlier, the scaled interaction energy
difference $\epsilon$
behaves, if the one-dimensional interaction energy 
$E_{N-1,1}^{1\text{d},\text{int}}$
is calculated using $g_{1\text{d}}$, very similarly for the 
$(1,1)$, $(2,1)$ and $(3,1)$ systems. This suggests
that the behavior of $\epsilon$ is governed by
two-body physics and that usage of 
$g_{1\text{d}}^{\text{trap}}$ instead
of $g_{1\text{d}}$ should lead to an improved one-dimensional treatment
for the $(2,1)$ and $(3,1)$ systems.
To corroborate this premise,
we consider the weakly-interacting regime.
Treating the Hamiltonian $H_{1\text{d}}$
for the $(2,1)$
system, see Eq.~(\ref{eq_ham1d}), 
with $g_{1\text{d}}$ replaced by $g_{1\text{d}}^{\text{trap}}$ 
in second-order perturbation theory
[the $(3,1)$ system can be treated analogously], we find
\begin{eqnarray}
\label{eq_pt21}
\frac{E_{2,1}^{1\text{d},\text{int}}}{\hbar \omega_{\rho}} \approx
d_1^{(1)}
\frac{g_{1\text{d}}^{\text{trap}}}{\hbar \omega_{\rho} a_{\rho}}
+
d_1^{(2)}
\left( 
\frac{g_{1\text{d}}^{\text{trap}}}{\hbar \omega_{\rho} a_{\rho}} \right)^2,
\end{eqnarray}
where $d_1^{(1)}=3 c_1^{(1)}/2$ and
\begin{eqnarray}
d_1^{(2)} = 
\frac{3}{8 \pi}\left[ -3 +2 \sqrt{3} + \ln (2+\sqrt{3}) - 4 \ln (2) \right]
 \approx \nonumber \\ -0.118355081.
\end{eqnarray}
The fact that $d_1^{(1)}$ is not equal
to the number of interacting pairs times $c_1^{(1)}$ 
can be interpreted as being a consequence of the
Pauli exclusion principle.
Rewriting Eq.~(\ref{eq_pt21}) as a series in $g_{1\text{d}}$, we find that the
coefficient of the linear term is unchanged while the
coefficient of the 
quadratic term becomes
$d_1^{(2)}+d_2^{(2)}$
with
$d_2^{(2)}=3 c_2^{(2)}/2$.
For $\eta=10$, this yields
$d_1^{(2)}+d_2^{(2)} \approx -0.12054$.
Since the change of the quadratic coefficient 
arises from the two-body coupling constant, it
is attributed to a two-body effect.

Ideally, we would perform an analogous perturbative treatment for the
Hamiltonian $H_{3\text{d}}$.
It turns out, however, that the calculations are
somewhat involved since the three-dimensional
second-order perturbation theory
sums need to be regulated~\cite{johnson1,johnson2}.
Thus, we instead
fit the full three-dimensional energies
$E_{2,1}^{3\text{d}}/(\hbar \omega_{\rho})$ to a power series in
$g_{1\text{d}}$. The fit yields the same linear coefficient
as the perturbative treatment of $H_{1\text{d}}$
and a slightly more negative coefficient 
for the quadratic term, $-0.120754251(1)$,
where the number in round brackets denotes the 
uncertainty of the fit (the uncertainty of the three-dimensional
energies is negligible for this analysis).

We attribute the small difference 
of $-0.00021$ between the quadratic coefficients of the
three- and one-dimensional energies 
to a three-body effect.
Specifically, the coefficient of the quadratic term
can be decomposed, following ideas
developed in Refs.~\cite{johnson1,johnson2}
for bosons under spherically symmetric harmonic confinement, 
into two parts, one that accounts for effective
two-body interactions and one that accounts for effective
three-body interactions.
The effective two-body interactions of the one- and three-dimensional
models 
should agree since the
one-dimensional analysis uses $g_{1\text{d}}^{\text{trap}}$.
The effective three-body interactions 
of the one- and three-dimensional
models, however, differ slightly since the three-dimensional
Hamiltonian has, compared to the one-dimensional
Hamiltonian, extra transverse modes (or virtual excitations) that
are available during collision processes.
Quantifying the effective three-body interactions
away from the weakly-interacting regime 
is beyond the scope of this paper. Our numerical
results suggest, though, that they are relatively small.
We note that calculations for the $(2,1)$
system in a harmonic  wave guide predict that the
inverse of the odd-channel atom-dimer
scattering length is proportional
to $g_{1\text{d}}^3$~\cite{mora05a} and 
$g_{1\text{d}}^4$~\cite{petrov}, respectively, in the
small $g_{1\text{d}}$ limit. These effects are of higher order than the
effective three-body interaction discussed above for the trapped system.

\begin{figure}
\centering
\vspace*{0.5in}
\includegraphics[angle=0,width=0.4\textwidth]{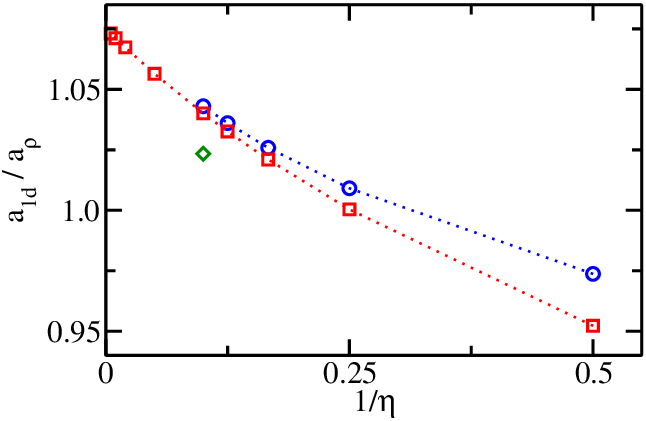}
\vspace*{0.5in}
\caption{(Color online)
Quantifying the deviations between the full three-dimensional and 
approximate one-dimensional treatment for the energetically
lowest lying molecular branch as a function of $1/\eta$.
The squares, circles and diamond show the scattering length 
$a_{1\text{d}}$ for which 
the full three-dimensional and the approximate one-dimensional
zero-range interaction energies deviate by 50\% 
for the $(1,1)$, $(2,1)$ and $(3,1)$ systems,
respectively.
For the $(1,1)$ and $(2,1)$ systems, dotted lines are shown as a 
guide to the eye.
 }\label{fig_energyeta}
\end{figure}

Next, we determine the dependence of $\epsilon$,
calculated using the one-dimensional Hamiltonian 
$H_{1\text{d}}$ with $g_{1\text{d}}$ [see Eq.~(\ref{eq_ham1d})],
on $\eta$ in the regime where $\epsilon$ 
$(\epsilon>0$) is not small compared to $1$.
The squares, circles and diamond in Fig.~\ref{fig_energyeta}
show the $a_{1\text{d}}/a_{\rho}$ value
for which $\epsilon$ is equal to $1/2$
as a function of $\eta$.
It can be seen that the $a_{1\text{d}}/a_{\rho}$
value depends only weakly on $\eta$. 
Moreover, for the impurity problem considered,
the $a_{1\text{d}}/a_{\rho}$
value depends only weakly on $N$.
If we look for the $a_{1\text{d}}/a_{\rho}$
values for which $\epsilon$ takes values different from $1/2$, we find
similar results [though
the ordering of the curves for the different $(N-1,1)$
systems depends on the specific $\epsilon$ value considered].
This implies that the accuracy of the effective 
one-dimensional treatment can be estimated quite reliably from the 
results for the two-body problem for a single $\eta$.
Intuitively, this can be understood from the fact that
the $(N-1,1)$ system can be thought of as consisting of a single 
diatomic molecule and $N-2$ unpaired atoms.

\begin{figure}
\vspace*{0.5in}
\centering
\includegraphics[angle=0,width=0.4\textwidth]{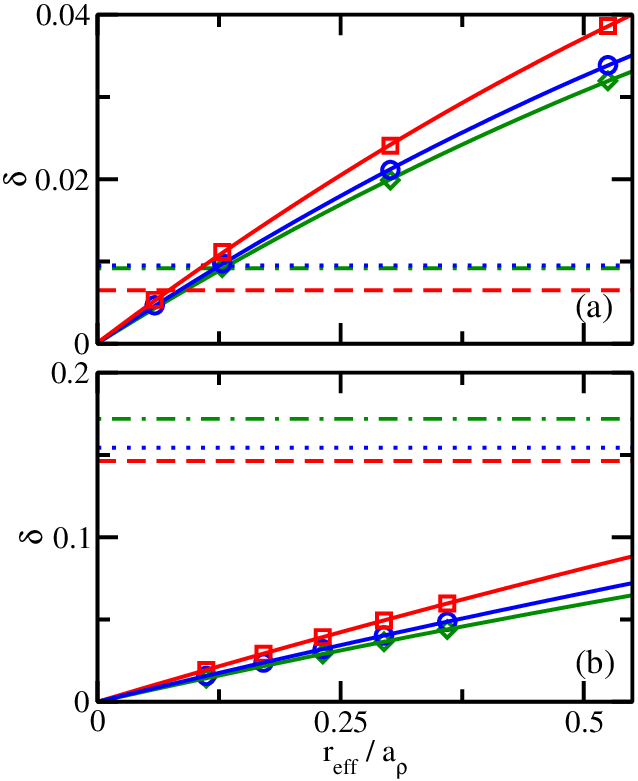}
\vspace*{0.5in}
\caption{(Color online)
Quantifying the finite-range effects of the full three-dimensional energies
for $\eta=10$.
(a) Squares, circles and diamonds show the scaled energy difference
$\delta$ at $a_{z}/a_{3\text{d}}=-10$ as a function of the effective
range $r_{\text{eff}}$ for the $(1,1)$,
$(2,1)$ and $(3,1)$ systems, respectively.
(b) Squares, circles and diamonds show the scaled energy difference
$\delta$ at $a_{z}/a_{3\text{d}}=-2$ as a function of the effective
range $r_{\text{eff}}$ for the $(1,1)$, 
$(2,1)$ and $(3,1)$ systems, respectively.
(a) and (b): Solid lines are three-parameter fits to the numerical data.
For comparison, dashed, dotted and dash-dotted 
lines show the scaled interaction energy 
difference
$\epsilon$ for the $(1,1)$, $(2,1)$ and $(3,1)$ systems, respectively.
}\label{fig_energyrange}
\end{figure}

Lastly, we 
estimate the dependence of the relative energies obtained
from the three-dimensional treatment  on the effective range $r_{\text{eff}}$,
which is defined through the low-energy expansion of the free-space
$s$-wave scattering phase shift $\delta_0(k)$,
$k \cot(\delta_0(k)) = -1/a_{3\text{d}}+r_{\text{eff}} k^2 /2$~\cite{newton}.
Here, $k$ denotes the relative scattering wave vector.
To quantify this dependence, we define
the scaled energy
difference $\delta$,
\begin{eqnarray}
\delta = \frac{E_{N-1,1}^{3\text{d,int}}(r_{\text{eff}}=0)-
E_{N-1,1}^{3\text{d,int}}(r_{\text{eff}} \ne 0)}{|E_{N-1,1}^{3\text{d,int}}(r_{\text{eff}}=0)|},
\end{eqnarray}
where the energies $E_{N-1,1}^{3\text{d,int}}(r_{\text{eff}}=0)$
and $E_{N-1,1}^{3\text{d,int}}(r_{\text{eff}} \ne 0)$ are the zero-range and finite-range
interaction energies, respectively.
Figures~\ref{fig_energyrange}(a) and \ref{fig_energyrange}(b)
show the quantity $\delta$ for 
$a_{z}/a_{3\text{d}}=-10$ 
[$g_{1\text{d}} / (\hbar \omega_{\rho} a_{\rho})\approx -0.477$]
and $a_{z}/a_{3\text{d}}=-2$
[$g_{1\text{d}} / (\hbar \omega_{\rho} a_{\rho})\approx -1.20$],
respectively.
Squares, circles and diamonds show the scaled energy difference
for the $(1,1)$, $(2,1)$ and $(3,1)$ systems and 
solid lines show fits to the numerical data.
To compare the finite-range effects with the 
difference between the three-dimensional and one-dimensional
energies,
dashed, dotted  and dash-dotted lines show the scaled interaction energy 
difference
$\epsilon$ for the $(1,1)$, $(2,1)$ and $(3,1)$ systems, respectively.
For two $^{6}$Li atoms, the van
der Waals length $r_{\text{vdW}}$ is 
$31.26 a_0$~\cite{ChinReview,Yan96}, where $a_0$ denotes the 
Bohr radius.
Using the values of the $s$-wave scattering
length and the van der Waals length~\cite{Gao98,Flambaum99},
we find---for the parameters of the Heidelberg 
experiment~\cite{Jochim3,thesisGerhard}---$r_{\text{eff}} \approx 0.0141 a_{\rho}$ for
$a_{z}/a_{3\text{d}}=-10$ 
and $r_{\text{eff}} \approx 0.0138 a_{\rho}$
for
$a_{z}/a_{3\text{d}}=-2$.
Inspection of Fig.~\ref{fig_energyrange} shows that the finite-range
corrections to the interaction energy are roughly a
factor of 10 and 100 smaller
for $a_{z}/a_{3\text{d}}=-10$ and 
$a_{z}/a_{3\text{d}}=-2$, respectively,
than the difference between the three- and one-dimensional energies.
This implies that the finite-range effects can, to a good
approximation, be neglected in analyzing cold atom experiments 
in highly elongated traps such as those
conducted by the Heidelberg group.

\section{Summary}
\label{sec_summary}
This paper discussed the energies of 
Fermi gases 
with a single impurity under highly-elongated confinement.
We presented energies for the $(1,1)$, $(2,1)$ and $(3,1)$
systems and assessed the accuracy 
of an effective one-dimensional Hamiltonian parametrized by the
effective one-dimensional coupling constant $g_{1\text{d}}$.
We focused on states that can be reached 
experimentally by 
first preparing an effectively non-interacting system and by then
adiabatically changing the $s$-wave scattering length through
application of an external magnetic field. 
As has been shown in the literature~\cite{BlumeReview}, 
the complete energy spectra 
of few-body systems are rather dense and exhibit avoided
crossings (if the states belong to the same subspace of the
full Hilbert space) and sharp crossings (if the states
belong to different subspaces of the full Hilbert space).

We found, in agreement with what might be expected naively, 
that the validity regime of the effective one-dimensional
treatment based on $H_{1\text{d}}$
[see Eq.~(\ref{eq_ham1d})] is limited, for the molecular branch,
to the regime where the one-dimensional
even parity scattering length is larger than the harmonic oscillator
length in the tight confinement direction.
When the one-dimensional even parity scattering length is large
compared to the harmonic oscillator
length in the tight confinement direction,
we found that the effective one-dimensional description can be improved 
if $g_{1\text{d}}$ is 
replaced by $g_{1\text{d}}^{\text{trap}}$, 
which explicitly accounts
for trap corrections that arise from the fact that the aspect ratio
$\eta$ is finite and not infinitely large. 
When the one-dimensional even parity scattering length is small
compared to the harmonic oscillator
length in the loose confinement direction,
we found that the effective one-dimensional description can be improved 
if the leading-order energy dependence
of $g_{1\text{d}}^{\text{trap}}$ is 
accounted for.
The fact that 
the leading-order energy dependence 
of the effective-one-dimensional coupling constant
can be derived from the waveguide Hamiltonian  (i.e., 
the Hamiltonian with $\omega_z=0$)
indicates that the physics in this regime is not unique to the trapped
system but analogous to what has been found for waveguide
geometries.
Moreover, our analysis suggests that the accuracy of the effective
one-dimensional treatment of small Fermi systems with an impurity
can be assessed fairly accurately
by looking at the $(1,1)$ system. 
Lastly, we found that finite-range effects (or equivalently, the
energy-dependence of the three-dimensional scattering length 
$a_{3\text{d}}$) are
negligible
for the conditions of the
Heidelberg experiment~\cite{Jochim3,thesisGerhard}.

Acknowledgement:
Support by the ARO and valuable discussions with Yangqian Yan
are gratefully acknowledged.

\onecolumngrid


\begin{thebibliography}{10}
\bibitem{explain_parity}
The parity $\Pi_z$ is positive (negative) if the wavefunction does not
(does) change sign under the operation that sends all $z_j$ to $-z_j$.
The parity $\Pi_{\vec{\rho}}$ is positive (negative) if the wavefunction 
does not
(does) change sign under the operation that sends all $x_j$ to $-x_j$
and all $y_j$ to $-y_j$.
\end{thebibliography}
\end{document}